# A NONLINEAR TRANSFORM BASED ANALOG VIDEO TRANSMISSION FRAMEWORK

*Yongtao Liu, Xiaopeng Fan, Member, IEEE, Yang Wang, Debin Zhao,*
*Member, IEEE and Wen Gao, Fellow, IEEE*

*Abstract*—SoftCast, a cross-layer design for wireless video transmission, is proposed to solve the drawbacks of digital video transmission: threshold effect and leveling-off effect. Since only linear transforms are used in SoftCast, in this paper, we propose a nonlinear transformed analog transmission framework achieving the same effect. Specifically, in encoder, we carry out power allocation on the transformed coefficients $X_{ij}^{1/a}$ and encode the coefficients based on the new formulation of power distortion. In decoder, the process of LLSE estimator is also improved. Accompanied with the inverse nonlinear transform, DCT coefficients can be recovered depending on the scaling factors $b_i$, LLSE estimator coefficients $w_i$ and metadata. Experiment results show that our proposed framework outperforms the SoftCast in PSNR 1.08 dB and the MSSIM gain reaches to 2.35% when transmitting under the same bandwidth and total power.

## 1. INTRODUCTION

Contemporary video communication frameworks are mainly divided into three categories: digital video coding, analog video coding and hybrid digital-analog video coding. Traditional digital video transmission system adopt separated source-channel coding framework. Video sequences are first compressed into bitstream through a standard video encoder, such as H.264/AVC [1]. Then the bitstream is encoded by a channel encoder before transmission. It is well-known that the separated source-channel design has two inherent drawbacks, called threshold effect and leveling-off effect [2]. The threshold effect means the receiver cannot decode the received bit steam when the channel is worse than a certain threshold and the leveling-off effect means that the receiver cannot reconstruct video at a quality matching with the channel SNR when the channel is better than expected. In this case, channel conditions have not been sufficiently used and the highest performance is determined in the encoder. In multi-user scenarios, it is hard to satisfy receivers with various channel conditions through broadcasting.

To ensure different receivers can get different reconstructed video matching with their channel, a cross-layer design named SoftCast [3], [4] has been proposed and it has obtained remarkable achievement. Unlike the conventional digital coding scheme, SoftCast adopts joint source-channel coding scheme. It uses discrete cosine transform (DCT) and power allocation complete the aim of compression and error protection. Fig. 1 shows the result of a group of pictures (GoP) before and after 3D-DCT. We can see that, result of natural pictures after 3D-DCT transform is high compact. So we can get a highly similar reconstruction with only a small part of the components. This is the theoretical basis of compressing in analog communication scheme.

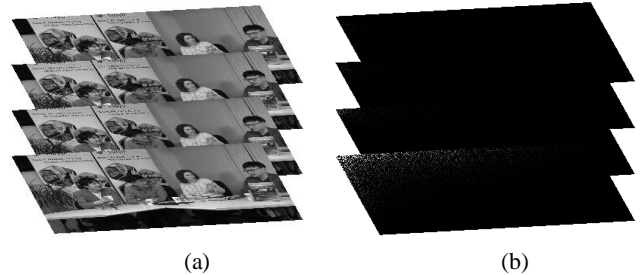

(a)          (b)

**Fig. 1.** (a) Original pictures and (b) DCT transform results.

SoftCast redistributes the power and bandwidth among DCT coefficients instead of the binary steams. If the bandwidth is no enough, the less important coefficients (i.e. coefficients with smaller variances) are dropped to satisfy the bandwidth capacity. Benefit from the novel design in SoftCast, channel perturbations are translated into approximation in the original video pixels and therefore the receiver reconstruct the video sequences at a quality commensurate with the channel condition while eliminating the threshold effect and level-off effect.

SoftCast performs gracefully while dealing with various channel conditions and only linear transforms are used. Based on this observation, it is possible to improve the same effect when using nonlinear transforms in analog transmission. In this paper, we propose a nonlinear transformed analog video transmission framework. The DCT coefficients are transformed with a nonlinear function and then we derive the new distortion formulation. Corresponding power allocation

Yongtao Liu, Xiaopeng Fan, Yang Wang and Debin Zhao are with Harbin Institute of Technology, China. Wen Gao is with Peking University, China. Xiaopeng Fan is the corresponding author of this paper (e-mail: fxp@hit.edu.cn).

is implemented among the transformed coefficients. That is the process before amplitude modulation at the encoder. The decoder use the linear least square estimation (LLSE) [5] and inverse power transform to change the received coefficients to the DCT coefficients and then pixel values are reconstructed. Experiment results shows that our proposed nonlinear transformed framework outperforms SoftCast both in PSNR and SSIM.

The remainder of the paper is organized as follows. Section 2 reviews the related work. Section 3 describes our proposed communication scheme. Experiment results are reported in Section 4 and Section 5 concludes the paper.

## 2. BACKGROUND

### 2.1. Related Work

Conventional digital video transmission scheme separates source coding and channel coding. Motion estimation, transformation, quantization and entropy coding are used to compress the data and increase the robustness. These techniques have been widely used in modern video coding standards, such as H.264/AVC [1] and HEVC [6]. However, the visual quality of compressed video is sensitive to the channel perturbation. To adapt to the various channel conditions, Choi *et al.* [16] realized adaptive coding by adopting different quantization parameters. Thomas *at el.* [8] proposed a scalable video coding (SVC) scheme, solving the level-off effect in a progressive way. In SVC, the coded streams are divided into one basic layer and several enhancement layers.

For analog video transmission, a novel design, SoftCast [3], has been proposed to eliminate the level-off effect and the threshold effect. Based on SoftCast, many works have been presented to improve the video quality and the compression ratio. Fan *et al.* [9] proposed a soft mobile video broadcast scheme based on distributed source coding (D-cast), applying distributed source coding to exploit the temporal redundancy. Wu *et al.* [10] explored the spatial correlation by applying coset coding across adjacent pixel lines. Xiong *at el.* [11] have verified that decorrelation transform can bring significant gain by boosting the energy diversity in the signal representation.

For hybrid video transmission, many hybrid schemes have been proposed to integrate the high efficiency of digital video transmission and the elegant performance of the analog video transmission. Liu *et al.* [12] proposed a hybrid scheme, in which the residuals were encoded by ParCast [13] and other parts were encoded with a digital encoder. Besides, Zhao *et al.* [14] proposed an adaptive hybrid digital–analog video transmission scheme (A-HDAVT), in which each GoP was filtered into one low-pass frame and several high-pass frames, transmitted with the digital transmission method and the analog transmission method respectively. Tan *et al.* [15] proposed a prediction model to optimize the resource allocation for a superposition coding based hybrid digital-analog system.

### 2.2. Review of SoftCast

SoftCast is a comprehensive design for wireless video broadcast, with the function of video compression, error protection and data transmission. As shown in Fig. 2, the encoder of SoftCast consists of DCT, power allocation, Walsh-Hadamard transform (WHT). The decoder consists of inverse WHT, LLSE, and inverse DCT.

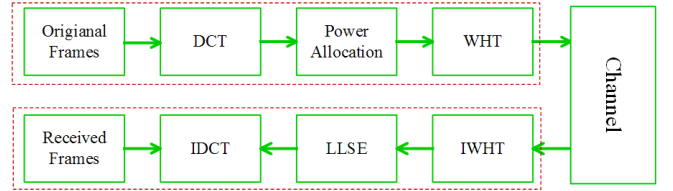

**Fig. 2.** Flow chart of SoftCast.

In encoder, first, DCT removes the spatial redundancy of a video frame. Then power allocation minimizes the total distortion by optimally scaling the DCT coefficients. WHT redistributes the energy among transmitted packets to protect the data from packets loss. Finally, before transmission, coded data are mapped to wireless symbols by quadrature amplitude modulation (QAM).

In decoder, coded data can be obtained after demodulation and inverse WHT. The LLSE estimator is used as inverse operation of power allocation and denoising. The overall process of encoding and decoding can be represented as follows:

$$Y_i = F_i(X_i) = g_i X_i$$
$$\check{Y}_i = Y_i + N$$
$$\hat{X}_i = G_i(\check{Y}_i) = \omega_i \check{Y}_i \quad (1)$$

where $X_i$ denote the coefficients in chunk $i$, $F_i$, $G_i$ represent the encoding process and decoding process respectively, $g_i$ is the scaling factor, $Y_i$ represent the encoded coefficients, $\check{Y}_i$ is the received data, $\omega_i$ is the LLSE factor and the $\hat{X}_i$ represents the decoding DCT coefficients.

Chunk division is processed before power allocation to satisfy the bandwidth. When the bandwidth is constrained, some chunks with non-zero values are discarded gradually. As distortion resulting from the discarded chunks is the sum of the squares of the coefficients, the chunks with smaller variances are more possible to be discarded.

## 3. PROPOSED FRAMEWORK

### 3.1. Framework Overview

Our proposed framework is shown in Fig. 3. First, each GoP is transformed with 3D-DCT and divided into chunks.

As most DCT coefficients are close to zero, containing little information of the original frames and non-zero coefficients are spatially clustered. The number of chunks transmitted is adaptive according to the bandwidth. Then we transform the DCT coefficients with a power function $f(x) = x^{1/a}$ and reallocate power among the transformed coefficients. WHT is used to balance the energy among transmitted packets. Fig. 4 shows the data distribution of a chunk before and after the power function. We can see that the transformed coefficients is more clustered comparing with the original coefficients. Regardless of the symbol, the encoding process with nonlinear transform can be expressed as

$$Y_i = F_i(X_i) = b_i X_i^{1/a}, \quad (2)$$

where the $X_i$ represents the DCT coefficient of chunk $i$, $1/a$ means the power of the power function, $b_i$ denotes the scale factors and $Y_i$ is the encode results.

In decoder, after the demodulation and inverse WHT, the received data can be expressed as

$$\breve{Y}_i = Y_i + N \quad (2)$$

where $N$ denotes the channel noise. The factors of LLSE estimator will be used to denoise the received data $\breve{Y}_i$. Therefore, DCT coefficients can be approximated as

$$\hat{X}_i = G_i(\breve{Y}_i) = \omega_i \breve{Y}_i^a \quad (3)$$

where $G_i(*)$ denotes the decoding function, $\omega_i$ is new the LLSE factor and $\hat{X}_i$ represents the decoding results. The frames can be reconstructed with inverse DCT by setting all the discarded chunks to zero.

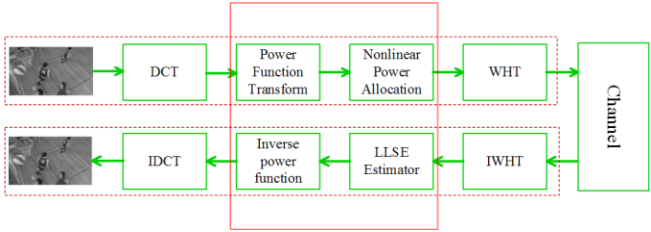

**Fig. 3.** Flow chart of our proposed framework.

### 3.1. Nonlinear Transform Based Distortion Optimization

Power allocation plays an important role in the analog video transmission schemes, which intends to minimize the total distortion within the constraint of total power. We first transform the coefficients with a nonlinear function and assign power among chunks of transformed coefficients. We derive the new formulation of the total distortion, which contains the distortion of SoftCast as a specific case. Related to SoftCast, we have higher degree of freedom of the representation the DCT coefficients. Nonlinear transform perform better than SoftCast in reallocating power within chunks.

Since power function is used in our framework, the decoding process for each chunk can be expressed as

$$\breve{Y}_i = \omega_i Y_i = \omega_i (b_i X_i^{\frac{1}{a}} + N)^a \quad (4)$$

then the distortion of chunk $i$

$$D_i = E\left(\left\| X_i - \omega_i \left(b_i X_i^{\frac{1}{a}} + N\right)^a \right\|^2\right) \quad (5)$$

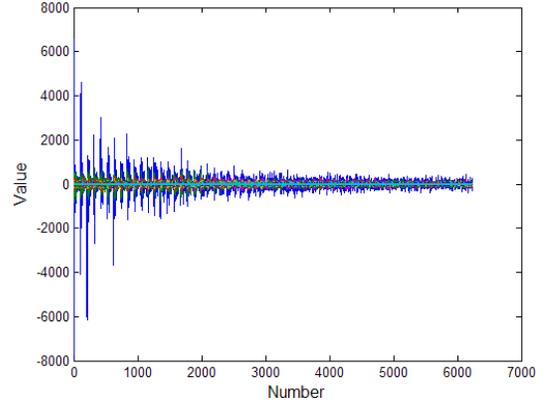

(a) Original DCT coefficients

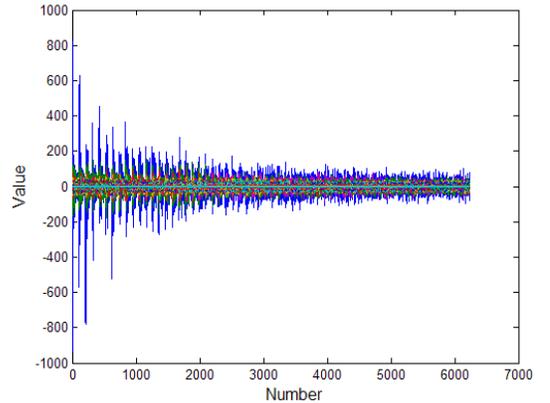

(b) Transformed DCT Coefficients

**Fig. 4.** Distribution of DCT coefficients before and after the power function transform.

#### 3.1.1. Optimization Formulation

In this paper, we model the original coefficients $X_i$ as random values with zero mean and variance $\sigma_{i0}^2$, transformed coefficients $X_i^{\frac{1}{a}}$ with zero mean and variance $\sigma_{i1}^2$, random variables $X_i^{1-\frac{1}{a}}$ with zero-mean and $\sigma_{i2}^2$ and the channel is an additive Gaussian noise channel with variance $\sigma_n^2$.

So the total distortion in the receiver with a constraint of total power P can be formulated as

$$\min \quad \sum_i D_i = \sum_i E(\left\| X_i - \left[\omega_i \left(b_i X_i^{\frac{1}{a}} + N\right)\right]^a \right\|^2) \quad (6)$$

$$s.t. \quad P = \sum_i b_i^2 \sigma_{i1}^2 \quad (7)$$

we use Taylor expansion to approximate formula (5) for convenience

$$D_i \approx E(\left\| X_i - \omega_i \left(b_i^a X_i - a b_i^{a-1} X_i^{1-\frac{1}{a}} N\right)\right\|^2) \quad (8)$$

assuming $\left[\omega_i \left(b_i X_i^{\frac{1}{a}} + N\right)\right]^a$ making a good approximation of $X_i$, so $b_i^a \omega_i$ approximate 1 in high SNR according to (8). $D_i$ can be rewritten as

$$D_i \approx E(\left\| a b_i^{-2} X_i^{1-\frac{1}{a}} N \right\|^2) \quad (9)$$

So the optimization problem with the constraint of total power P can be simply expressed in the form of variances as

$$\min \quad \sum_i D_i = \sum_i a^2 \sigma_{i2}^2 \frac{\sigma_n^2}{b_i^2} \quad (10)$$

$$s.t. \quad P = \sum_i b_i^2 \sigma_{i1}^2 \quad (11)$$

*3.1.2. Lagrange Multiplier*

We use Lagrange multiplier to solve the optimization problem of (10) and (11), since $a^2$ and $\sigma_n^2$ are constant, the Lagrange function $L(\alpha, b_1, ..., b_M)$ can be simplified as

$$L(\alpha, b_1, ..., b_M) \approx \sum_i \frac{\sigma_{i2}^2}{b_i^2} - \alpha \left(P - \sum_i b_i^2 \sigma_{i1}^2\right) \quad (12)$$

making $\frac{\partial L_1}{b_i} = 0, \frac{\partial L_1}{\alpha} = 0$, we can get

$$\alpha = \frac{\sum \sigma_{i1} \sigma_{i2}}{P} \quad (13)$$

the nonlinear encoder that minimize the distortion is

$$Y_i = b_i X_i^{1/a}, where$$

$$b_i = \frac{1}{\sqrt{\sigma_{i1}}} \sqrt{\frac{P \sigma_{i2}}{\sum \sigma_{i2} \sigma_{i1}}} \quad (14)$$

Where the $\sigma_{i1}, \sigma_{i2}$ represents the standard deviation of $X_i^{\frac{1}{a}}$ and $X_i^{1-\frac{1}{a}}$.

### 3.2. LLSE Estimator

Accompanied with the encoded video data, a small amount of metadata are also transmitted to receiver for decoding. In our framework, we also need to transmit the mean of each chunk, variances of $X_i, X_i^{1-1/a}$ and $X_i^{1/a}$, namely $\sigma_{i0}^2, \sigma_{i2}^2$ and $\sigma_{i1}^2$. Besides, a bitmap recording the location of transmitted chunks also need to be sent to the receiver.

According to the new formulation of total distortion, we recalculate the LLSE estimator factors for denoising. We get a LLSE coefficient $\omega_i$ for each chunk, which is related to $\sigma_{i0}^2, a, \sigma_{i2}^2, \sigma_n^2$ and scaling factor $b_i$. After getting the approximation of the nonlinear transformed coefficients, inverse power allocation will be adopt to obtain DCT coefficients.

At the receiver, we know the encode coefficients with noise of each chunk after inverse WHT. LLSE can be represented in a simple form as

$$\check{Y}_i = \omega_i \, (b_i X_i^{\frac{1}{a}} + N)^a \quad (15)$$

Similar to the process of distortion optimization, the total distortion in the principle of minimize mean-square error (MMSE) can be formulated as

$$D = \sum_i E(\left\| X_i - \left[\omega_i \left(b_i X_i^{\frac{1}{a}} + N\right)\right]^a \right\|^2)$$

$$\approx \sum_i ((1 - b_i^a \omega_i)^2 \sigma_{i0}^2 + a^2 b_i^{2a-2} \omega_i^2 \sigma_{i2}^2 \sigma_n^2) \quad (16)$$

Obviously that $D$ is a convex function of variables $\omega_i$ for the other variables are constants in decoder. Distortion achieve the global minima when all the partial derivatives of $D$ equal to zero and the LLSE estimator factors

$$\omega_i = \frac{\sigma_{i0}^2}{b_i^a (\sigma_{i0}^2 + a^2 \sigma_{i2}^2 \frac{\sigma_n^2}{b_i^2})} \quad (17)$$

The distortion $D$ can be calculated by putting $\omega_i$ back into the formula

$$D = \sum \frac{a^2 \sigma_{i2}^2 \sigma_{i0}^2 \sigma_n^2}{b_i^2 (\sigma_{i0}^2 + a^2 \sigma_{i2}^2 \frac{\sigma_n^2}{b_i^2})} \quad (18)$$

## 4. EXPERIMENT RESULTS

The test platform of the experiments is MATLAB R2014a. Test videos in this paper are in the common test condition of HEVC. In this paper, values of $a$ are 1.11 and 1.12, 1.31, 1.20 and 1.29 for videos with different resolutions empirically. To evaluate the performance of the proposed method for different constraints of the channel, the signal noise ratio (SNR) is set 5, 10, 15, and 20. PSNR and SSIM are used as the metrics.

Table 1 shows the PSNR error between our proposed framework and SoftCast. The average gain can reach to 0.47 dB, 0.73 dB, 0.94 dB and 1.08 dB when SNR values 20, 15

10 and 5 respectively. The corresponding maximum can reach to 3.0 dB, 3.7 dB, 4.0 dB and 4.2 dB respectively. The extreme values present in the video, 'SlideShow'. We analyzed the video and found that most frames contain less contents relative to the other videos. The nonlinear transform analog video transmission framework execute better power allocation for smooth pictures.

**Table 1.** PSNR error between our proposed and SoftCast

| SNR | 20 | 15 | 10 | 5 |
|---|---|---|---|---|
| BasketballPass_416x240 | 0.0300 | 0.0463 | 0.0803 | 0.1293 |
| BlowingBubbles_416x240 | 0.1364 | 0.1946 | 0.2646 | 0.2853 |
| BQSquare_416x240 | 0.0791 | 0.1436 | 0.1534 | 0.1656 |
| RaceHorses_416x240 | 0.0743 | 0.1077 | 0.1212 | 0.1615 |
| BasketballDrill_832x480 | 0.0676 | 0.1236 | 0.1869 | 0.2323 |
| BQMall_832x480 | 0.1662 | 0.2672 | 0.3286 | 0.3675 |
| PartyScene_832x480 | 0.0850 | 0.1151 | 0.1395 | 0.1836 |
| RaceHorsesC_832x480 | 0.2000 | 0.2588 | 0.3003 | 0.3294 |
| FourPeople_1280x720 | 0.7203 | 1.1549 | 1.4721 | 1.6557 |
| Johnny_1280x720 | 0.8601 | 1.4866 | 1.9970 | 2.2772 |
| SlideEditing_1280x720 | 0.5934 | 0.6410 | 0.7022 | 0.7542 |
| SlideShow_1280x720 | 3.0006 | 3.7052 | 4.0278 | 4.1871 |
| BasketballDrive_1920x1080 | 0.3862 | 0.8455 | 1.3387 | 1.5987 |
| BQTerrace_1920x1080 | 0.3960 | 0.6294 | 0.8192 | 0.9746 |
| Cactus_1920x1080 | 0.4003 | 0.7606 | 1.0862 | 1.3572 |
| Kimono_1920x1080 | 0.5616 | 1.1246 | 1.6847 | 2.0164 |
| ParkScene_1920x1080 | 0.1633 | 0.3472 | 0.5750 | 0.7985 |
| Tennis_1920x1080 | 0.7018 | 1.4014 | 2.0792 | 2.4678 |
| PeopleOnStreet_2560x1600 | 0.3628 | 0.5354 | 0.6433 | 0.7265 |
| Traffic_2560x1600 | 0.3696 | 0.6274 | 0.8169 | 0.9479 |
| Average | 0.4677 | 0.7258 | 0.9409 | 1.0808 |

**Table 2.** MSSIM error between our proposed and SoftCast

| SNR | 20 | 15 | 10 | 5 |
|---|---|---|---|---|
| BasketballPass_416x240 | 0.0001 | 0.0002 | 0.0004 | 0.0009 |
| BlowingBubbles_416x240 | 0.0002 | 0.0006 | 0.0016 | 0.0040 |
| BQSquare_416x240 | 0.0004 | 0.0005 | 0.0018 | 0.0013 |
| RaceHorses_416x240 | 0.0003 | 0.0007 | 0.0013 | 0.0030 |
| BasketballDrill_832x480 | 0.0002 | 0.0007 | 0.0017 | 0.0034 |
| BQMall_832x480 | 0.0006 | 0.0016 | 0.0037 | 0.0065 |
| PartyScene_832x480 | 0.0001 | 0.0004 | 0.0010 | 0.0025 |
| RaceHorsesC_832x480 | 0.0005 | 0.0015 | 0.0034 | 0.0055 |
| FourPeople_1280x720 | 0.0026 | 0.0073 | 0.0174 | 0.0339 |
| Johnny_1280x720 | 0.0032 | 0.0092 | 0.0227 | 0.0464 |
| SlideEditing_1280x720 | 0.0054 | 0.0105 | 0.0147 | 0.0155 |
| SlideShow_1280x720 | 0.0555 | 0.0978 | 0.1338 | 0.1450 |
| BasketballDrive_1920x1080 | 0.0029 | 0.0084 | 0.0201 | 0.0340 |
| BQTerrace_1920x1080 | 0.0014 | 0.0041 | 0.0097 | 0.0177 |
| Cactus_1920x1080 | 0.0015 | 0.0044 | 0.0110 | 0.0224 |
| Kimono_1920x1080 | 0.0020 | 0.0060 | 0.0160 | 0.0367 |
| ParkScene_1920x1080 | 0.0005 | 0.0015 | 0.0043 | 0.0103 |
| Tennis_1920x1080 | 0.0029 | 0.0084 | 0.0218 | 0.0471 |
| PeopleOnStreet_2560x1600 | 0.0014 | 0.0041 | 0.0094 | 0.0170 |
| Traffic_2560x1600 | 0.0013 | 0.0036 | 0.0088 | 0.0172 |
| Average | 0.0041 | 0.0086 | 0.0152 | 0.0235 |

Experiment results also show that our proposed framework can outperform SoftCast slightly in lower resolution and perform better while dealing with higher resolution videos. Relative to SoftCast, our work also acquire better power allocation within the chunks. Fig. 6 show the parts of frame reconstruction contrast of the SoftCast and our proposed framework of video "Slideshow" and "Jonney".

Structural similarity index metrics (SSIM) [16] is an objective quality assessment method measuring videos and pictures quality and MSSIM [17] places more emphasis on partial structures. MSSIM is used to verify the performance of our proposed framework.

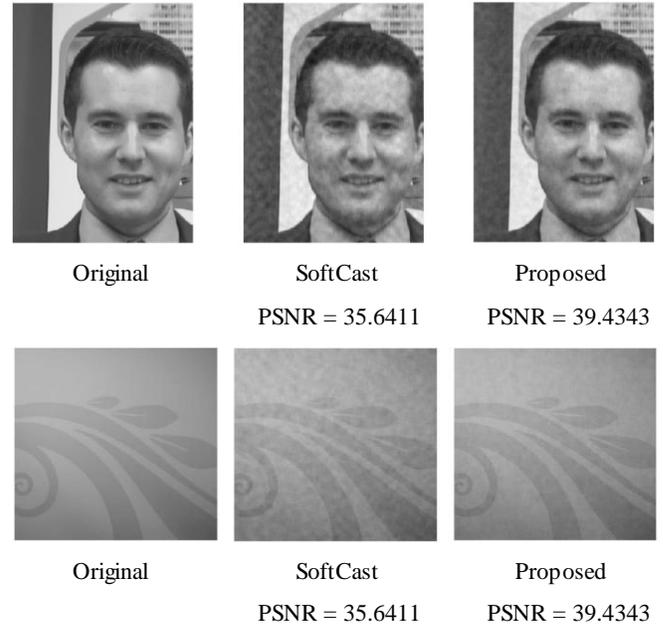

Original     SoftCast     Proposed

PSNR = 35.6411     PSNR = 39.4343

Original     SoftCast     Proposed

PSNR = 35.6411     PSNR = 39.4343

**Fig. 6.** MSSIM comparison of SoftCast and our work.

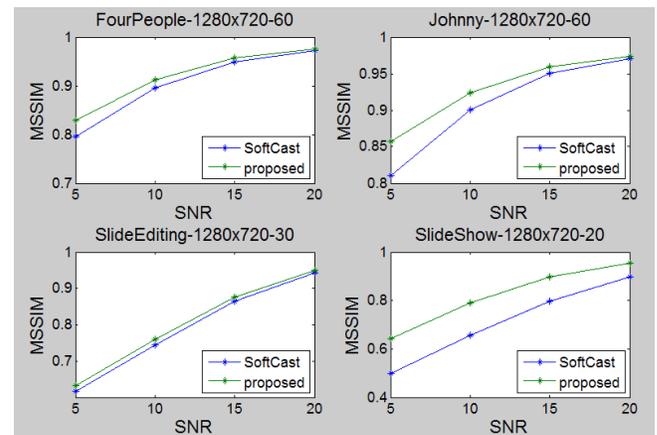

**Fig. 7.** MSSIM comparison of SoftCast and our framework

Table 2 shows the comparison of MSSIM between our proposed framework and the SoftCast. In our proposed framework, results show that MSSIM gain is positively related to PSNR gain. The average of MSSIM gain is 0.41%, 0.86%, 1.52% and 2.35% respectively. Fig. 7 shows some detail results of MSSIM contrast. Results confirm that our proposed framework can improve the performance of the analog video transmission both in PSNR and MSSIM.

## 5. CONCLUSION

In this paper, we propose a nonlinear transformed based analog video transmission framework. We execute power allocation on nonlinear transformed DCT coefficients instead of DCT coefficients itself and derive corresponding distortion expression with the constraint of total power. Scaling factors and LLSE estimator factors are updated by minimizing the distortion. Experiment result confirm that our proposed framework can improve the quality of reconstructed videos in PSNR and SSIM. In our future work, we will further expand the forms of the encode function and try to optimize power allocation inside the chunk instead of only allocating power among chunks.